\documentclass[11pt]{article}
\usepackage{graphicx, caption, subfigure}
\makeatletter
\newcommand{\singlespacing}{\let\CS=\@currsize\renewcommand{\baselinestretch}{1.0}\tiny\CS}
\newcommand{\doublespacing}{\let\CS=\@currsize\renewcommand{\baselinestretch}{1.5}\tiny\CS}

\begin{document}
\title{Ansatzs, Assumptions
and Production of J/$\Psi$-Particles: A Non-`Charmed' Approach vs.
the `Charmed' Ones}
\author{P.
Guptaroy$^1$\thanks{e-mail: gpradeepta@rediffmail.com(Communicating
author)}, Goutam Sau$^2$ \& S. Bhattacharyya$^3$\\
{\small $^1$ Department of Physics, Raghunathpur College,}\\
 {\small P.O.: Raghunathpur 723133,  Dist.: Purulia(WB), India.}\\
{\small $^2$ Beramara Ram Chandrapur High School,}\\
 {\small South 24-Parganas, 743609(WB), India.}\\
 {\small $^3$ Physics
and Applied Mathematics Unit(PAMU),}\\
 {\small Indian Statistical Institute,}\\
 {\small 203 B. T. Road, Kolkata - 700108, India.}}
\date{}
\maketitle
\bigskip
\bigskip
\begin{abstract}
We would attempt, in this work, at dwelling upon some crucial
aspects of $J/\Psi$-production in a few high energy nuclear
collisions in the light of a non-standard model which is outlined in
the text. The underlying physical ideas, assumptions and ansatzs
have also been enunciated in some detail. It is found that the
results arrived at with this main working approach here are fairly
in good agreement with both the measured data and the results
obtained on the basis of some other models of the `standard'
variety. Impact and implications of this comparative study have also
been precisely highlighted in the end.
\end{abstract}

 \newpage
 The proposed `suppression' of $J/\Psi$ mesons in the high energy
relativistic heavy ion collisions has been viewed althrough as one
of the most prominent diagnostics for formation of quark-gluon
plasma (QGP), the hypothetical `hot and dense matter' created in
high energy collisions \cite{matsui}-\cite{xu}. The RHIC experiments
could not yet give any final confirmation to the formation of any
clear and unambiguous quark-gluon plasma . Very recently, the ATLAS
group had reported the first results of $J/\Psi$ production at the
LHC in $Pb+Pb$ collisions \cite{atlas10}. Though the results are
still in rudimentary level, they could not give any decisive view in
regard to formation of deconfined matter. Still, both the
speculations about and searches on production-characteristics of
$J/\Psi$ continue to exist with much excitement and enthusiasm.
\par
We define our objectives here as: 1) to explain the main and major
features of the latest data on $J/\Psi$-production in BNL-RHIC
experiments from the proposed alternative approach called Sequential
Chain Model (SCM), built up by us in a set of previous works done in
the both the remote and recent past \cite{pgr10}; and 2) to compare
our model-based calculations with some other competing models.
\par
 According to this model, called
Sequential Chain Model (SCM),  high energy hadronic interactions
boil down, essentially, to the pion-pion interactions; as the
protons are conceived in this model as
$p$=($\pi^+$$\pi^0$$\vartheta$) \cite{pgr10}, where $\vartheta$ is a
spectator particle needed for the dynamical generation of quantum
numbers of the nucleons. The multiple production of $J/\Psi$-mesons
in a high energy proton-proton collisions is described in the
following way. The secondary $\pi$-meson or the exchanged
$\varrho$-meson emit a free $\omega$-meson and pi-meson; the pions
so produced at high energies could liberate another pair of free
$\varrho$ and trapped $\omega$-mesons (in the multiple production
chain). These so-called free $\varrho$ and $\omega$-mesons decay
quite  fast into photons and these photons decay into $\Psi$ or
$\Psi'$ particles, which, according to this alternative approach is
a bound state of $\Omega \bar{\Omega}$ or $\Omega' \bar{\Omega}'$
particles. In this model, then, obviously there is no concept on
parton fragmentation and recombination as in the standard model(SM).
Particles are emitted as just particles with their attributed
quantum numbers. Thus the ideas on 'evolution' do not arise in this
approach.
\par
The inclusive cross-section of the $\Psi$-meson produced in the $pp$
collisions given by \cite{pgr05}
\begin{equation}\displaystyle{
E\frac{d^3\sigma}{{dp}^3}|_{p+p\rightarrow{{J/\Psi}+X}}  \cong
C_{J/\Psi}\frac{1}{p_T^{N_R}}\exp(\frac{-5.35(p_T^2+m^2_{J/\Psi})}{<n_{J/\Psi}>^2_{pp}(1-x)})
\exp(-1.923{<n_{J/\Psi}>_{pp}}x),}
\end{equation}
where the expression for for average multiplicity for $\Psi$-particles in
$pp$ scattering would be given by
\begin{equation}\displaystyle{
<n_{J/\Psi}>_{pp} ~~~ = ~~~ 4\times10^{-6}s^{1/4}.}
\end{equation}
In the above expression, the term $|C_{J/\Psi}|$ is a
normalisation parameter and is assumed here to have a value $\cong
0.09$ for Intersecting Storage Ring(ISR) energy, and it is
different for different energy and for various collisions. The
terms $p_T$, $x$ and $m_{J/\Psi}$ represent the transverse
momentum, Feynman Scaling variable and the rest mass of the
$J/\Psi$ particle respectively. Moreover, by definition, $x ~ = ~
2p_L/{\sqrt s}$ where $p_L$ is the longitudinal momentum of the
particle. The $s$ in equation (2) is the square of the c.m. energy.
\par
The second term in the right hand side of the equation (1), the
constituent rearrangement term arises out of the partonic
rearrangements inside the proton. It is established that
hadrons (baryons and mesons) are composed of few partons. In the high energy interaction processes the
partons at large
transverse momenta undergo some dissipation losses due to the impact and
impulse of the projectile on the target and the parton inside them
(both the projectile and the target) they suffer some forced shifts
of their placements or configurations. These rearrangements mean
undesirable loss of energy , in so far as the production mechanism
is concerned. The choice of ${N_R}$ would depend on the following
factors: (i) the specificities of the interacting projectile and
target, (ii) the particularities of the secondaries emitted from a
specific hadronic or nuclear interaction and (iii) the magnitudes of
the momentum transfers and of a phase factor (with a maximum value
of unity) in the rearrangement process in any collision. The parametrisation is to be
done for two physical points, viz., the amount of momentum transfer
and the contributions from a phase factor arising out of the
rearrangement of the constituent partons. Assorting and combining
all these, we propose the relation to be given by \cite{pgr072}
\begin{equation}\displaystyle
N_R=4<N_{part}>^{1/3}\theta,
\end{equation}
where $<N_{part}>$ denotes the average number of participating
nucleons and $\theta$ values are to be obtained phenomenologically
from the fits to the data-points \cite{bhat882}.
\par
In order to study a nuclear interaction of the type $A+B\rightarrow
Q+ x$, where $A$ and $B$ are projectile and target nucleus
respectively, and $Q$ is the detected particle which, in the present
case, would be $J/\Psi$-mesons, the SCM has been adapted, on the
basis of the prescription by Wong \cite{wong}, to the Glauber
techniques by using Wood-Saxon distributions \cite{gorenstein}. The
inclusive cross-sections for $J/\Psi$ production in different
nuclear interactions of the types $A+B\rightarrow J/\Psi+ X$ in the
light of this modified Sequential Chain Model (SCM) can then be
written in the most generalised form as:
\begin{equation}\displaystyle
{E{\frac{d^3\sigma}{dp^3}}|_{A+B\rightarrow{J/\Psi}+ X}=
a_{J/\Psi} {p_T}^{-N_R} \exp(-c(p_T^2+m^2_{J/\Psi}))
\exp(-1.923{<n_{J/\Psi}>_{pp}x)}}.
\end{equation}
 where $a_{J/\Psi}$, $N_R$ and $c$ are
the factors to be calculated under certain physical constraints. The
set of relations to be used for evaluating the parameters
$a_{J/\Psi}$  are obtained from Wong \cite{wong}.
\par
As the psi-productions are generically treated rightly as the
resonance particles, the standard practice is to express the
measured $J/\Psi$ (total) crosssections times branching ratio to
muon or electrons, i.e.for lepton pairs , that is by
$B_{ll'}\sigma^{J/\Psi}_{p+p}$ .
\par
By using expression (4) we arrive at the expressions for the
differential cross-sections for the production of $J/\Psi$-mesons in
the mid and forward-rapidities (i.e. $|y|<0.35$ and $1.2<|y|<2.2$
respectively) in $p+p$ collisions at $\sqrt{s_{NN}}$=200 GeV at
RHIC.
\begin{equation}\displaystyle{
\frac{1}{2\pi p_T} B_{ll'} \frac{d^2\sigma}{dp_Tdy}|_{p+p\rightarrow
J/\Psi+X} = 6.1 p_T^{-1.183} \exp[-0.13(p_T^2+9.61)]
~~~~ for ~~|y|<0.35,}
\end{equation}
and
\begin{equation}\displaystyle{
\frac{1}{2\pi p_T} B_{ll'} \frac{d^2\sigma}{dp_Tdy}|_{p+p\rightarrow
J/\Psi+X} = 6.5 p_T^{-1.183} \exp[-0.16(p_T^2+9.61)]
~~~~ for ~~1.2<|y|<2.2.}
\end{equation}
For deriving the expressions (5) and (6) we have used the relation
$x\simeq \frac{2p_{Zcm}}{\sqrt s}= \frac{2m_T \sinh y_{cm}}{\sqrt
s}$, where $m_T$ , $y_{cm}$ are the transverse mass of the produced
particles and the rapidity distributions. $m_{J/\Psi} \simeq
3096.9\pm 0.011 MeV$ \cite{pdg} and $B_{ll'}$, the branching ratio
is for muons or electrons i.e. its for lepton pairs $J/\Psi
\rightarrow \mu^+\mu^-/e^+e^-$, is taken as $5.93\pm
0.10\times10^{-2}$ \cite{pdg} in calculating the above equations.
These expressions assume slightly altered numerical values for LHC
energy of 7 TeV.
\par
In Fig. 1(a) and Fig. 1(b), we have drawn the solid lines depicting
the SCM model-based results with the help of above four expressions
(5), (6) and some other changed forms of them against the
experimental measurements \cite{adare}, \cite{alice} respectively.
\par
For the calculation of the rapidity distribution from the set of
equations (1), (2), (3) and (4) we can make use of a standard
relation as given below:
\begin{equation}\displaystyle{
\frac{dN}{dy}=\int \frac{1}{2\pi p_T}\frac{d^2N}{dp_Tdy}dp_T}
\end{equation}
The rapidity distributions for the $J/\Psi$-production has now been
reduced to a simple relation given hereunder
\begin{equation}\displaystyle{
\frac{dN}{dy}={a_1}\exp(-0.23\sinh{y_{cm}}).}
\end{equation}
The normalization factor $a_1$ depends on the centrality of the
collisions and is obvious from the nature of the eqn.(1), eqn. (2),
eqn.(3) eqn.(4) and eqn. (7).
\par
For $p+p$ collisions, the calculated rapidity
distribution equation is given by
\begin{equation}\displaystyle{
\frac{dN}{dy}|_{p+p\rightarrow
J/\Psi+X}=1.215\times10^{-6}\exp(-0.23\sinh{y_{cm}}),}
\end{equation}
In Fig. 2 we have plotted the rapidity distributions for
$J/\Psi$-production in $p+p$ collisions.
Data in the figure are taken from Ref. \cite{adare4} and the
line shows the SCM-based output.
\par
From the expression (4), we arrive at the invariant yields for the
$J/\Psi$-production in $d+Au\rightarrow J/\Psi+X$ reactions for mid
and forward-rapidities.
\begin{equation}\displaystyle
\frac{1}{2\pi p_T}  \frac{d^2N}{dp_Tdy}|_{d+Au\rightarrow J/\Psi+X}
= 7.25\times10^{-7} p_T^{-0.629} \exp[-0.13(p_T^2+9.61)]~~~~ for ~~|y|<0.35,
\end{equation}
and
\begin{equation}\displaystyle
\frac{1}{2\pi p_T}  \frac{d^2N}{dp_Tdy}|_{d+Au\rightarrow J/\Psi+X}
= 4.25\times10^{-7} p_T^{-0.629} \exp[-0.16(p_T^2+9.61)]~~~~ for ~~1.2<|y|<2.2.
\end{equation}
In Figure 3, we have drawn the solid lines depicting the SCM-based
results with the help of equations (10) and (11) against the
experimental background \cite{adare4}.
\par
For $d+Au$ collisions, the calculated rapidity
distribution equation in the light of SCM turns into a form given below:
\begin{equation}\displaystyle{
\frac{dN}{dy}|_{d+Au\rightarrow
J/\Psi+X}=7.025\times10^{-6}\exp(-0.23\sinh{y_{cm}}),}
\end{equation}
In Fig. 4 we have plotted the rapidity distributions for
$J/\Psi$-production in $d+Au$ collisions. Data in the figure 4 are
taken from Ref.\cite{adare4}  and the solid line shows the SCM-based
theoretical outputs.
\par
Exactly, similar looking expressions have been worked out for
$Au+Au$ interactions with numerically changed physical parameters
depending on the energy differences and the mass number differences
of the target and the projectile.
\par
To calculate the rapidity distribution for $Au+Au$ collisions we are
taking into account of Eqs. (4) and (7). The rapidity distribution
for the 0-20$\%$ central $Au+Au$ collision is given by the following
equation
\begin{equation}\displaystyle{
\frac{dN}{dy}|_{Au+Au\rightarrow
J/\Psi+X}=0.472\times10^{-6}\exp(-0.23\sinh{y_{cm}}),}
\end{equation}
\par
Similarly, the solid lines in the Fig. 7 depict the theoretical plot
based on SCM (eqn. (13)) of $dN/dy$ vs. $y$ for 0-20$\%$ centrality
while the data for $Au+Au$ collisions are taken from
Ref.\cite{adare4}. The dotted line in the same figure shows
Coalescence Model-based result\cite{kahana}.
\par
Now we deal with nuclear modification factor (NMF) for production of
$J/\Psi$ in some nuclear collisions. Based on the standard
definition of NMF and the use of eqns. (9) and (12) of this work,
the expression of NMF finally turns out to be \cite{adare4}
\begin{equation}\displaystyle{
R_{AA}=\frac{dN^{AA}/dy}{<N_{coll}(b)>dN^{pp}/dy}.}
\end{equation}
For numerical calculations one has to use from Adare et al.
\cite{adare4}: $<N_{coll}(b)>_{dAu} \approx 15.1 \pm 1.0$ and
$<N_{coll}(b)>_{AuAu}\approx 955.4\pm 93.6$ .
\par
The plots shown in Fig.5 to Fig.8 depicting the results are almost
self-explanatory obviously from the figure-captions attached
thereto. Comparisons of our SCM-based results with two other
model-dependent calculations show neither sharp disagreement with
any of them, nor very good agreement with either of them which are
generically of standard model variety. Rather, our results are in
better agreement with data than either of them. So, in our opinion,
this work essentially represents a case of paradigm shift in the
domain of particle theory, as we have eschewed the conventional
views of $c \bar c$ approach to $J/\Psi$ in the `standard'
framework.
\par
Finally, we now summarise the conclusions: (a) The ``plasma" state
or the ``hot and dense" state is not to be any startling revelation,
because when heated to very high temperatures attained by the
extremely energetic collisions, the microscopic matter might be
converted to a liquid of somewhat unknown nature, and thus obviously
of a ``new" kind \cite{pgr10}; (b) Our treatment of the problem
rests only on hadron degrees of freedom. (c) The psion-production is
neither suppressed nor enhanced; rather in the present scenario this
is both qualitatively and quantitatively just natural, as in the
case with some other hadrons like pion, kaon etc.
\par
$\bf{ACKNOWLEDGEMENTS}$\\
The authors are grateful to the learned Referees for their helpful
comments and some very pointed, pertinent queries.
 
\begin{figure}
\subfigure[]{
\begin{minipage}{.5\textwidth}
\centering
\includegraphics[width=2.5in]{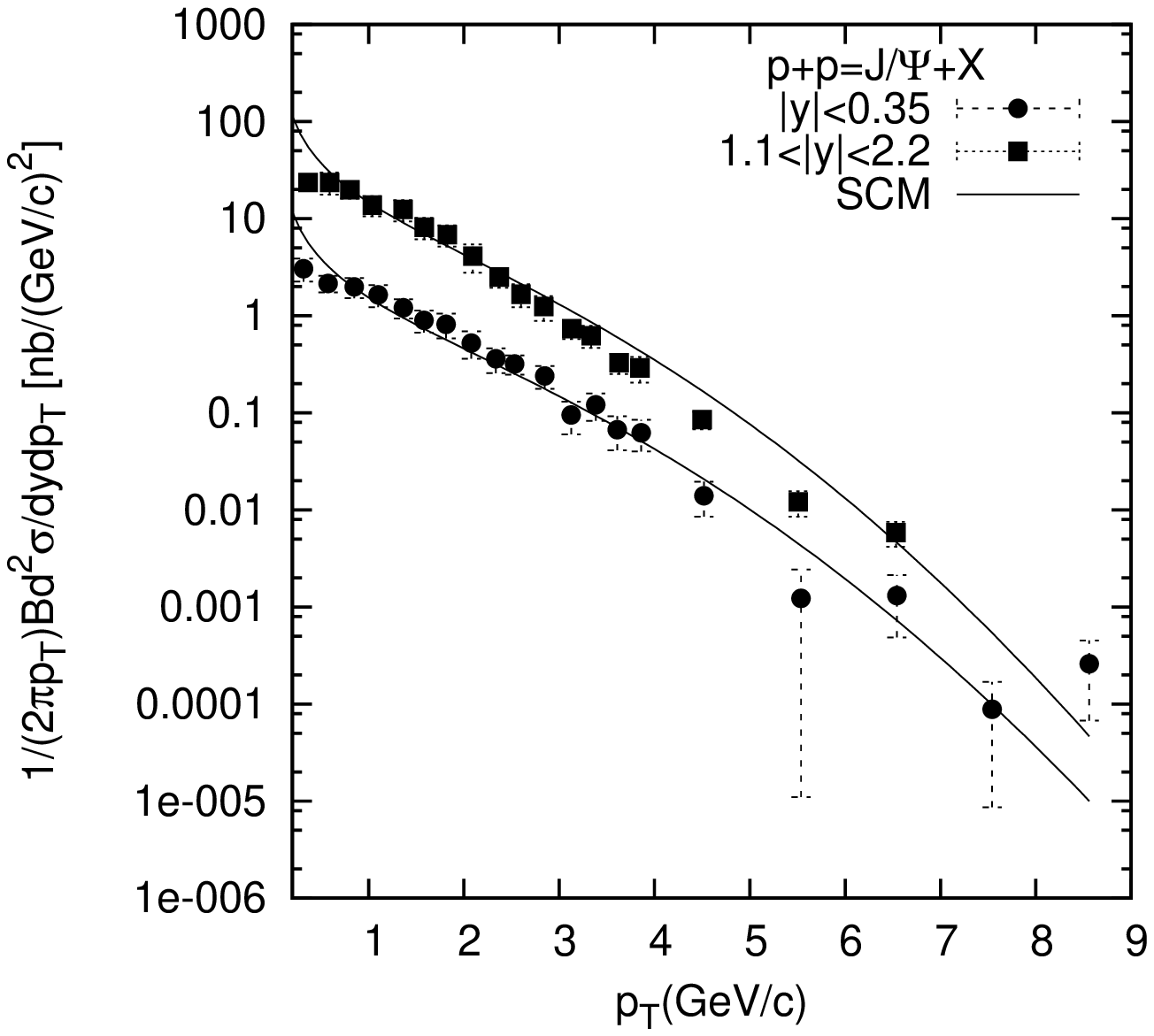}
\setcaptionwidth{2.6in}
\end{minipage}}%
\subfigure[]{
\begin{minipage}{0.5\textwidth}
\centering
 \includegraphics[width=2.5in]{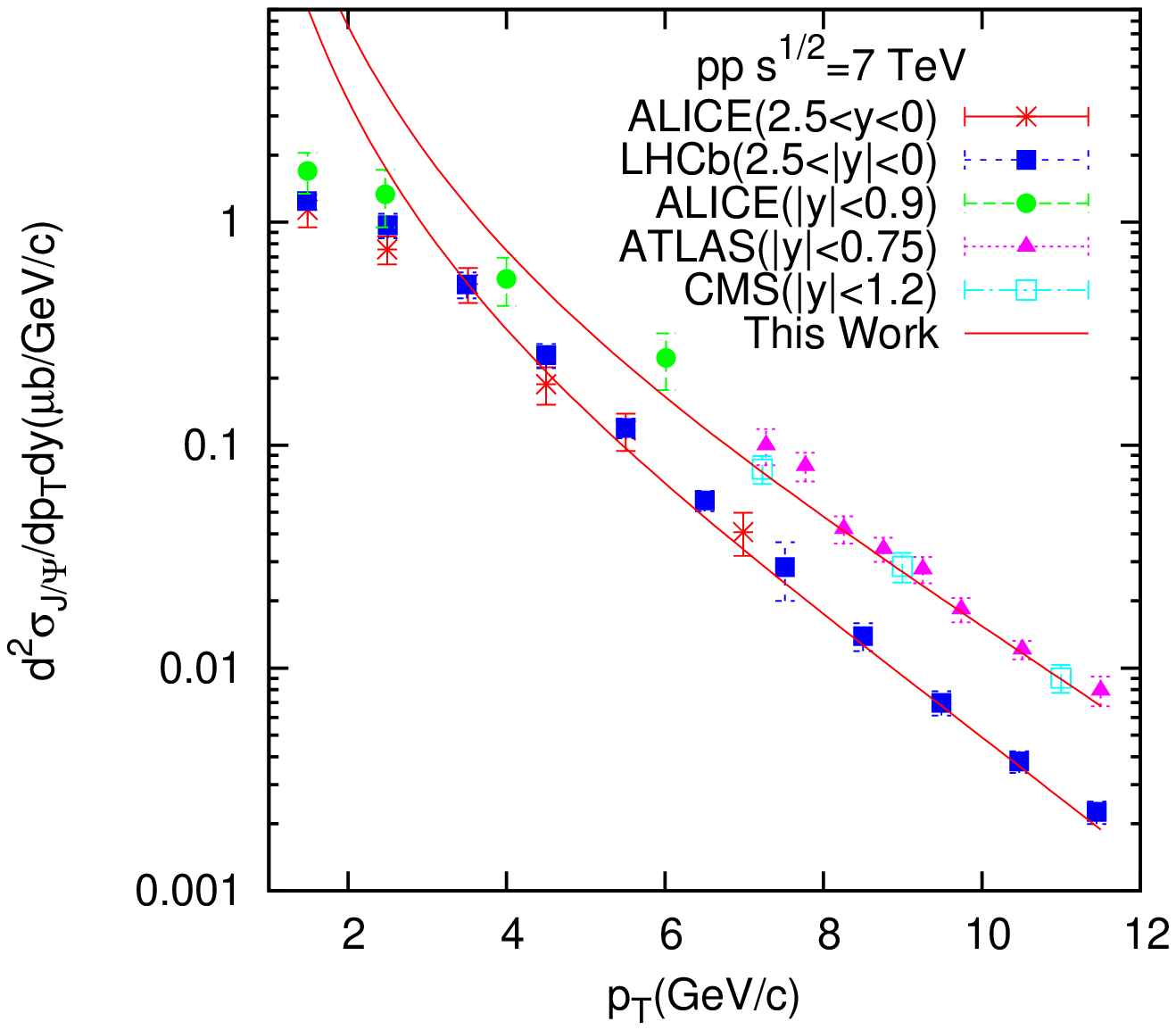}
  \end{minipage}}%
\caption{Plot of the invariant cross-section for $J/\Psi$
production in proton-proton collisions at (a) $\sqrt s_{NN} =200 GeV$ and (b) $\sqrt s_{NN} =7 TeV$ as
function of $p_T$. The data points are from \cite{adare} for (a) and from \cite{alice} for (b). The solid
curves show the SCM-based results.}
\end{figure}
\begin{figure}
\centering
\subfigure{
\begin{minipage}{.244\textwidth}
\centering
\includegraphics[width=1.8in]{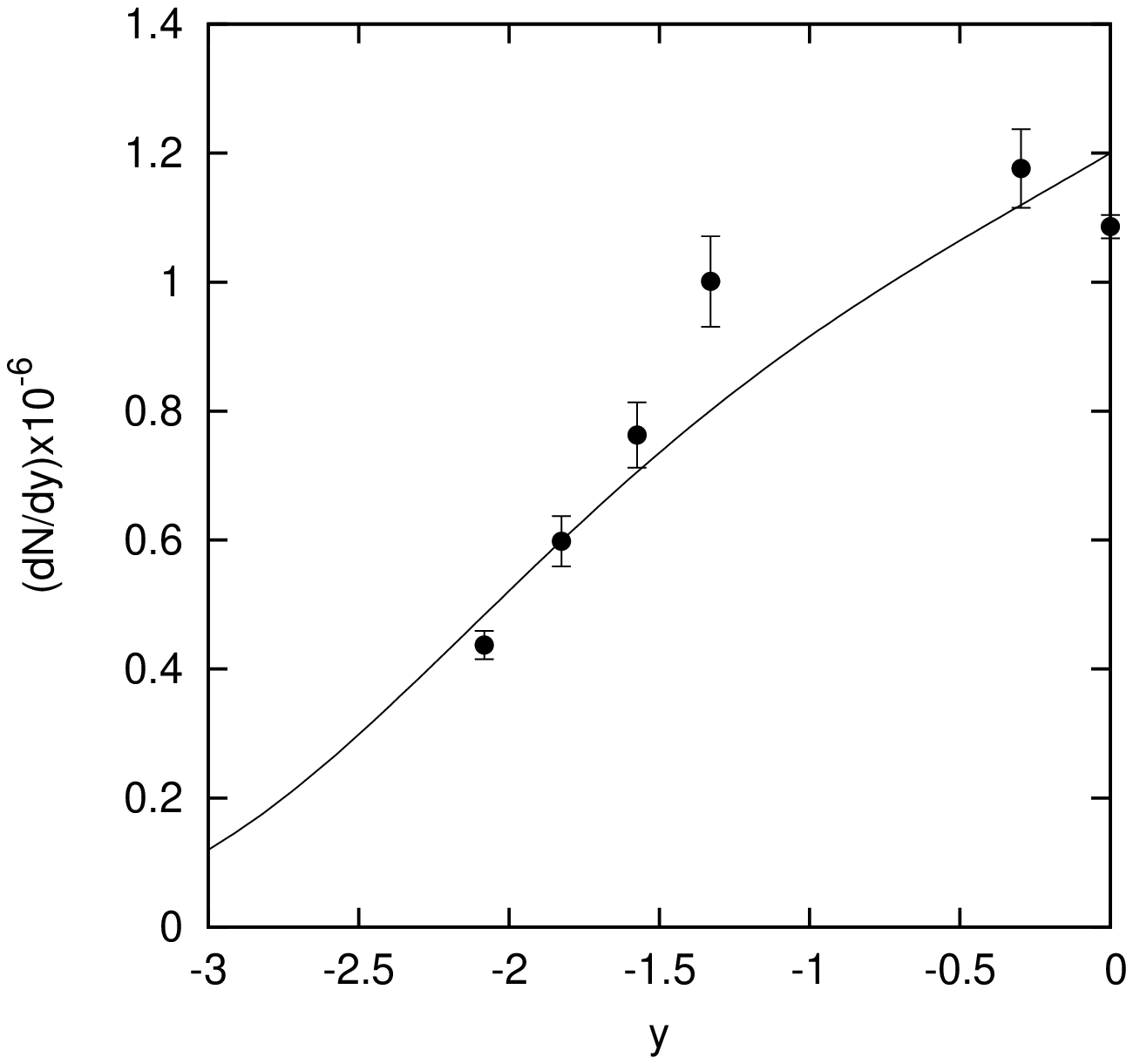}
\end{minipage}}%
\subfigure{
\begin{minipage}{0.24\textwidth}
\centering
 \includegraphics[width=1.8in]{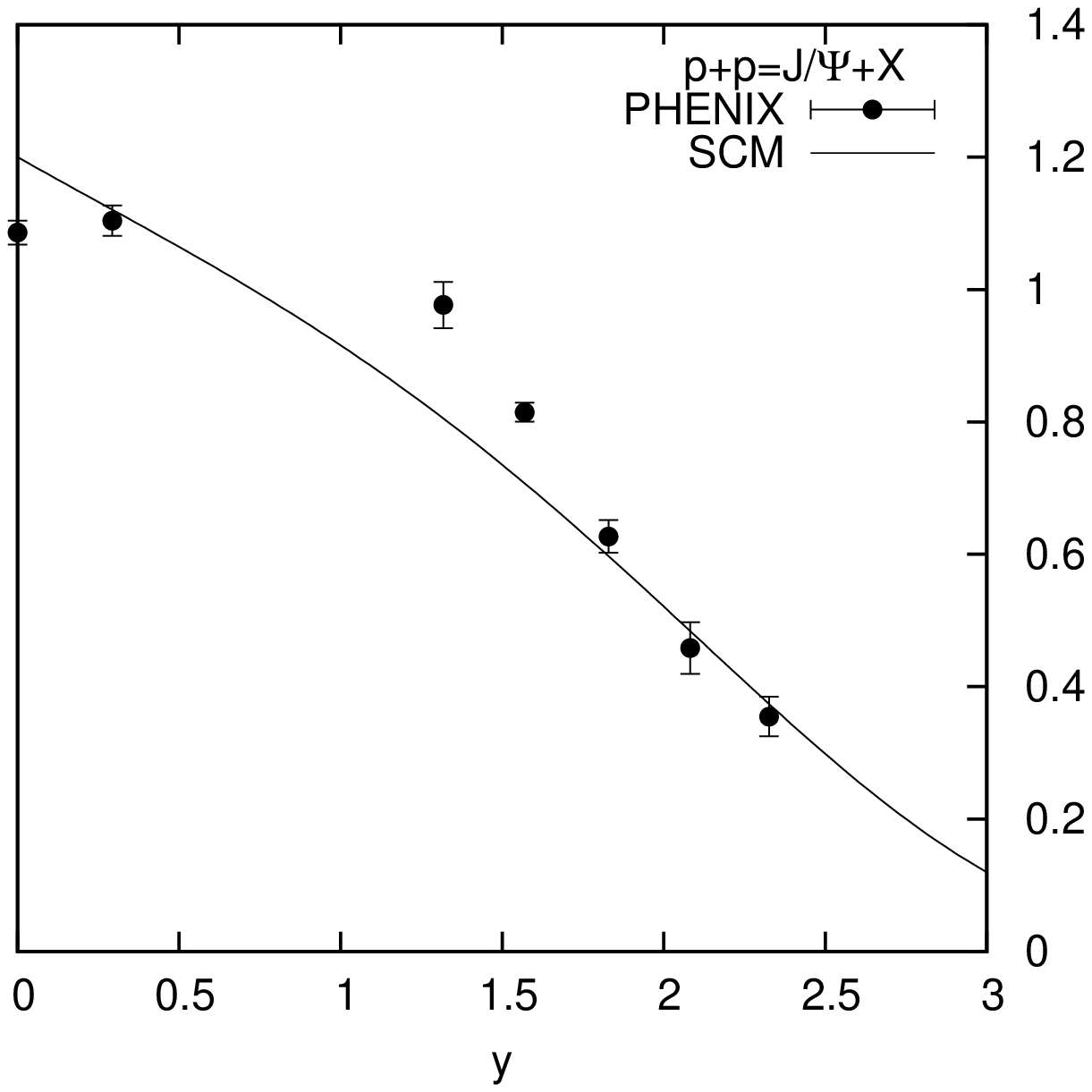}
  \end{minipage}}%
   \caption{ Plot of the rapidity distribution for $J/\Psi$ production in proton-proton
collisions at $\sqrt s_{NN} =200 GeV$ as function of $y$. The data
points are from \cite{adare4}. The solid curves show the SCM-based
results. }
\end{figure}
\begin{figure}
\centering
\includegraphics[width=2.5in]{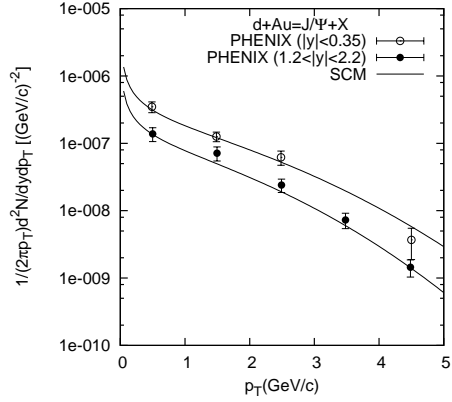}
\caption{Plot of the invariant yields for $J/\Psi$
production in $d+Au$ collisions at $\sqrt s_{NN} =200 GeV$ as
function of $p_T$. The data points are from \cite{adare4}. The solid
curves show the SCM-based results.}
\end{figure}
\begin{figure}
\centering
\subfigure{
\begin{minipage}{.244\textwidth}
\centering
\includegraphics[width=1.8in]{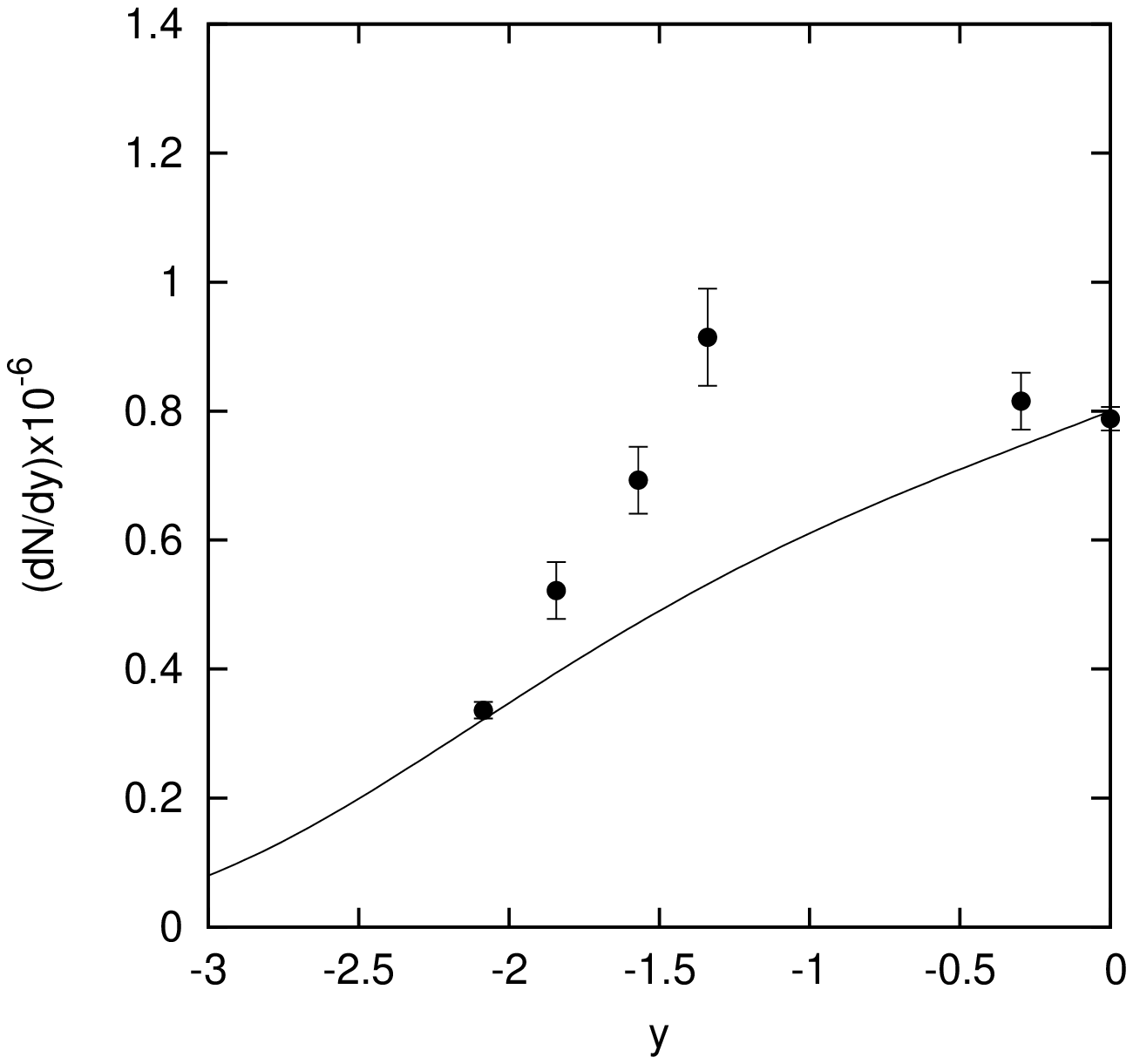}
\end{minipage}}%
\subfigure{
\begin{minipage}{0.24\textwidth}
\centering
 \includegraphics[width=1.8in]{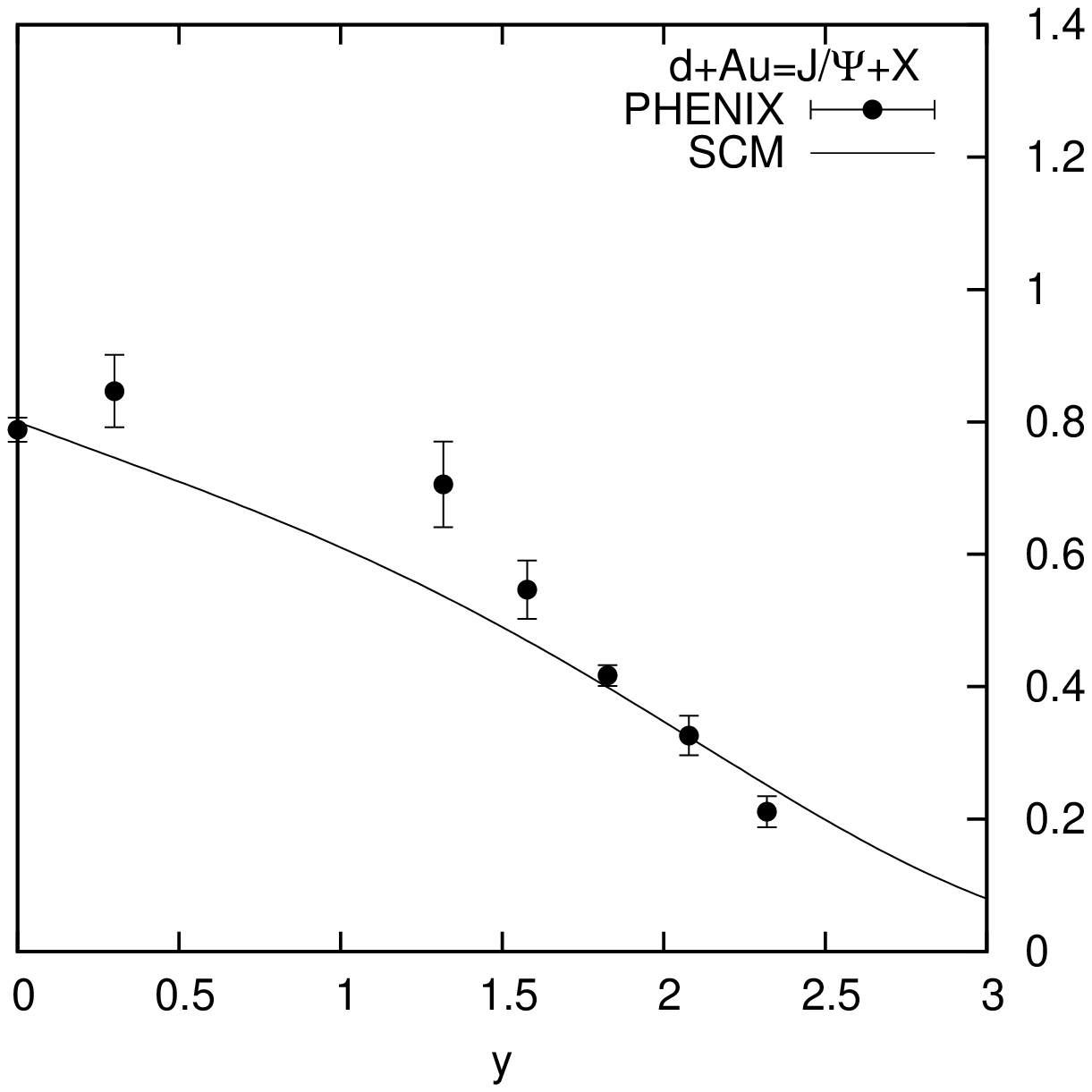}
  \end{minipage}}%
   \caption{Plot of the rapidity distribution for $J/\Psi$ production in $d+Au$
collisions at $\sqrt s_{NN} =200 GeV$ as function of $y$. The data
points are from \cite{adare4}. The solid curves show the SCM-based
results.}
\end{figure}
\begin{figure}
\centering
\includegraphics[width=2.5in]{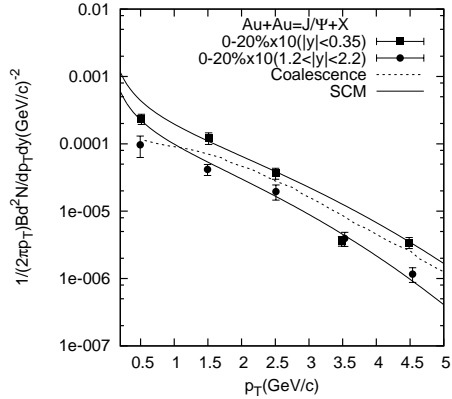}
\caption{Plot of the invariant yields for $J/\Psi$
production in $Au+Au$ collisions at $\sqrt s_{NN} =200 GeV$ as
function of $p_T$. The data points are from \cite{adare4}. The solid
curve shows the SCM-based results while the dotted curve depicts the Coalescence Model \cite{kahana}.}
\end{figure}
\begin{figure}
\centering
\subfigure{
\begin{minipage}{.244\textwidth}
\centering
\includegraphics[width=1.8in]{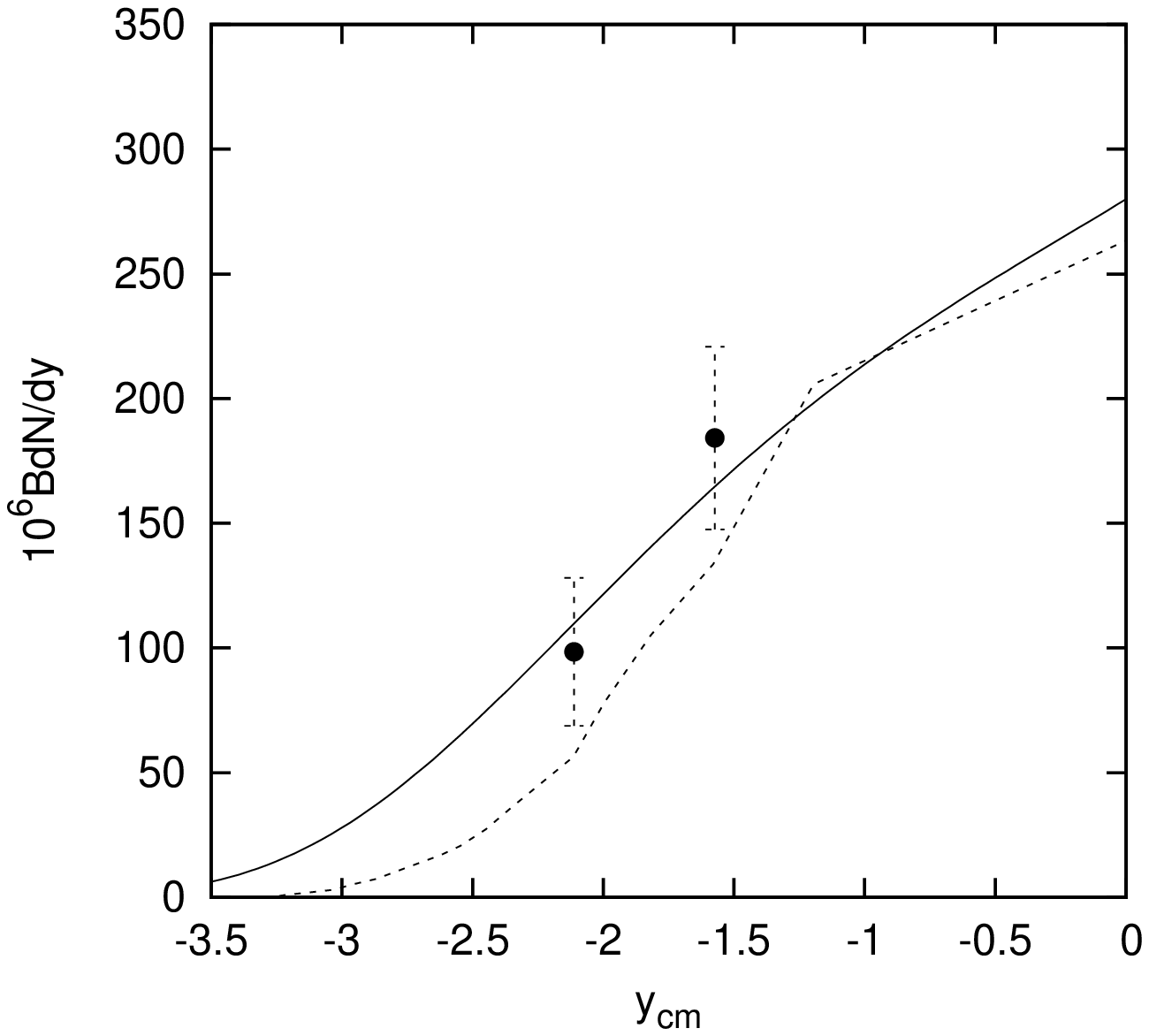}
\end{minipage}}%
\subfigure{
\begin{minipage}{0.24\textwidth}
\centering
 \includegraphics[width=1.8in]{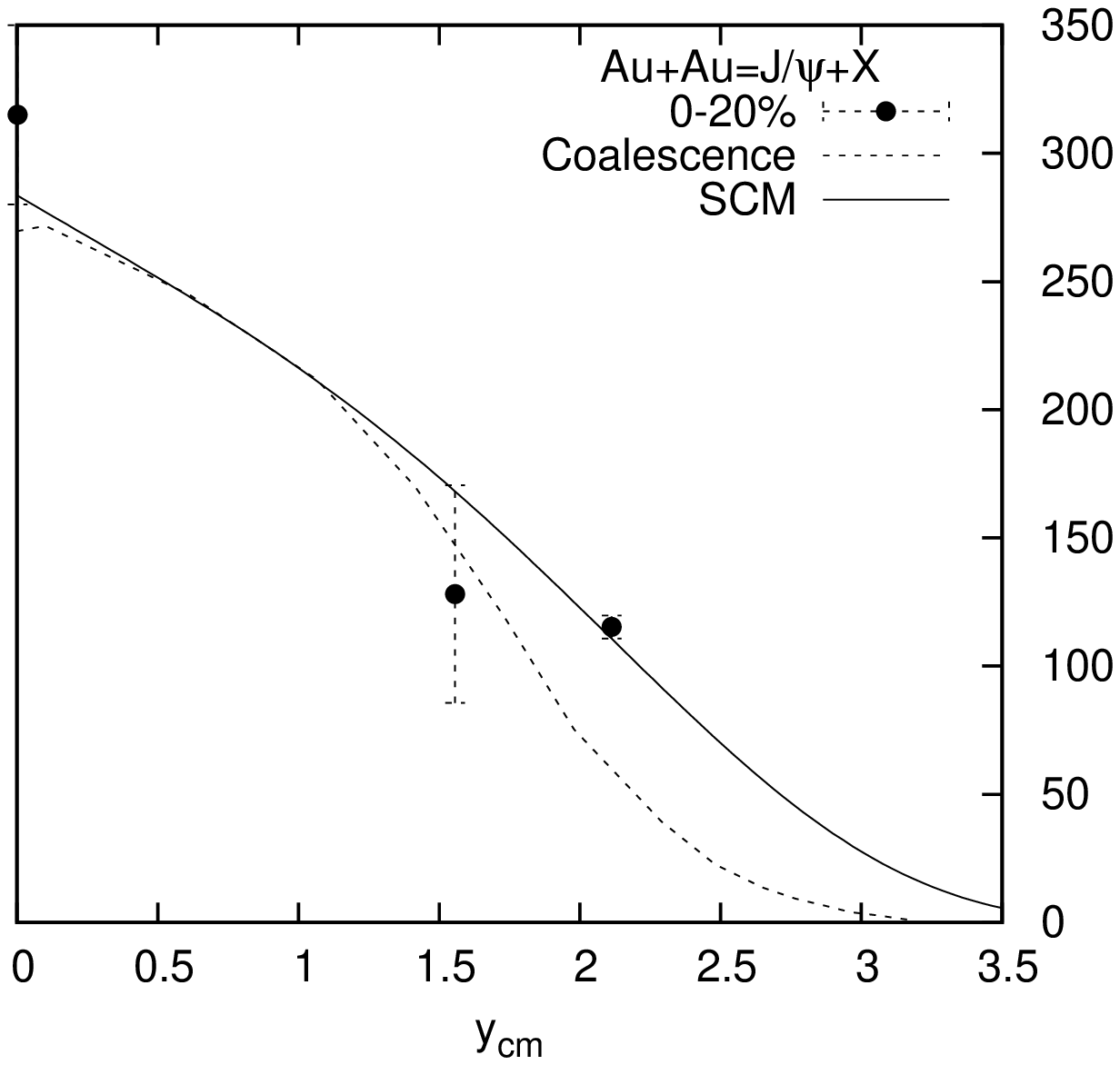}
  \end{minipage}}%
   \caption{Plot of the rapidity distribution versus $y_{cm}$ for $J/\Psi$ production in $Au+Au$
collisions at $\sqrt s_{NN} =200 GeV$. The data
points are from \cite{adare4}. The solid and dotted curves show respectively the SCM and the
Coalescence-oriented \cite{kahana}
results.}
\end{figure}
\begin{figure}
\subfigure[]{
\begin{minipage}{.5\textwidth}
\centering
\includegraphics[width=2.5in]{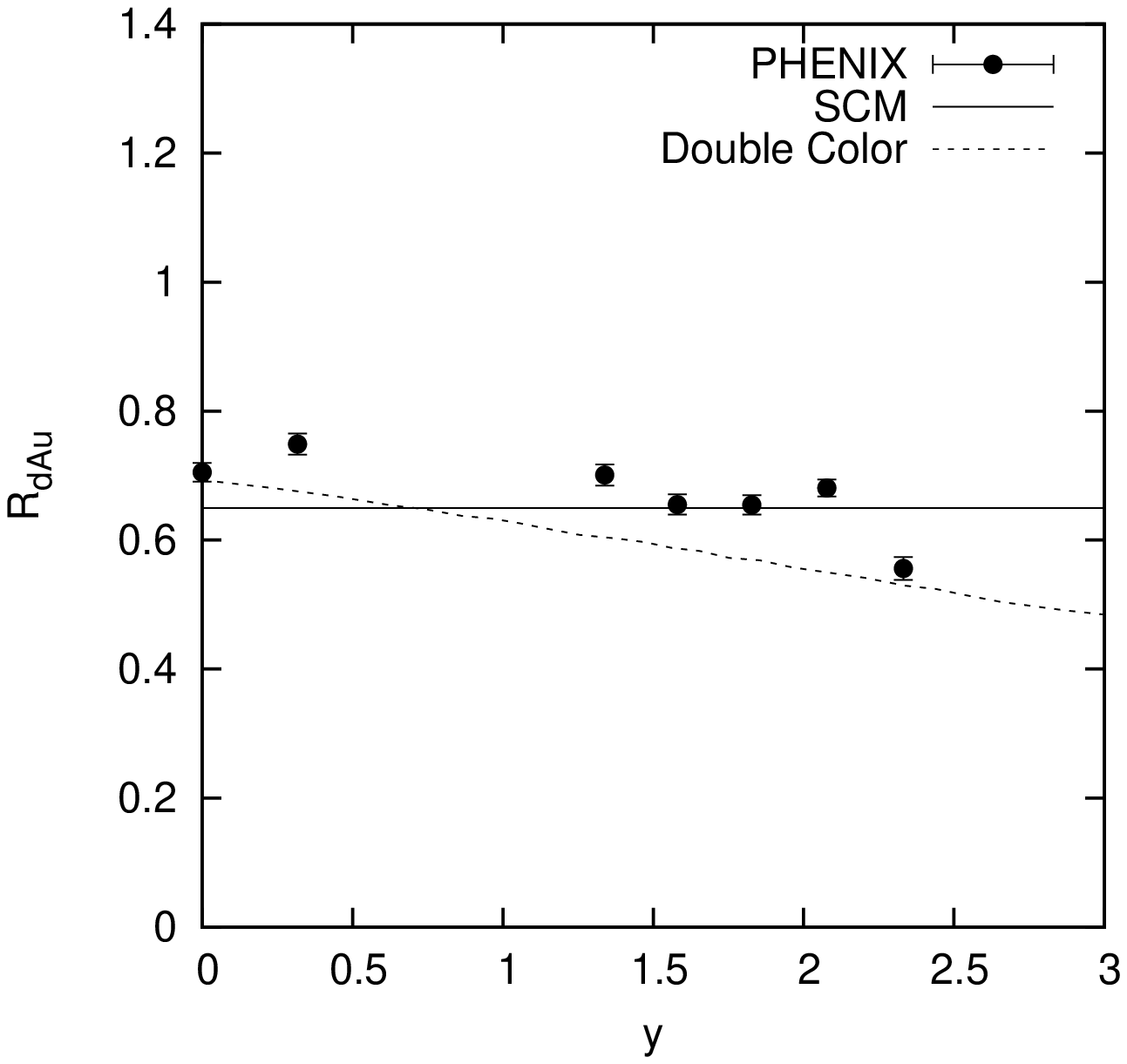}
\setcaptionwidth{2.6in}
\end{minipage}}%
\subfigure[]{
\begin{minipage}{0.5\textwidth}
\centering
 \includegraphics[width=2.5in]{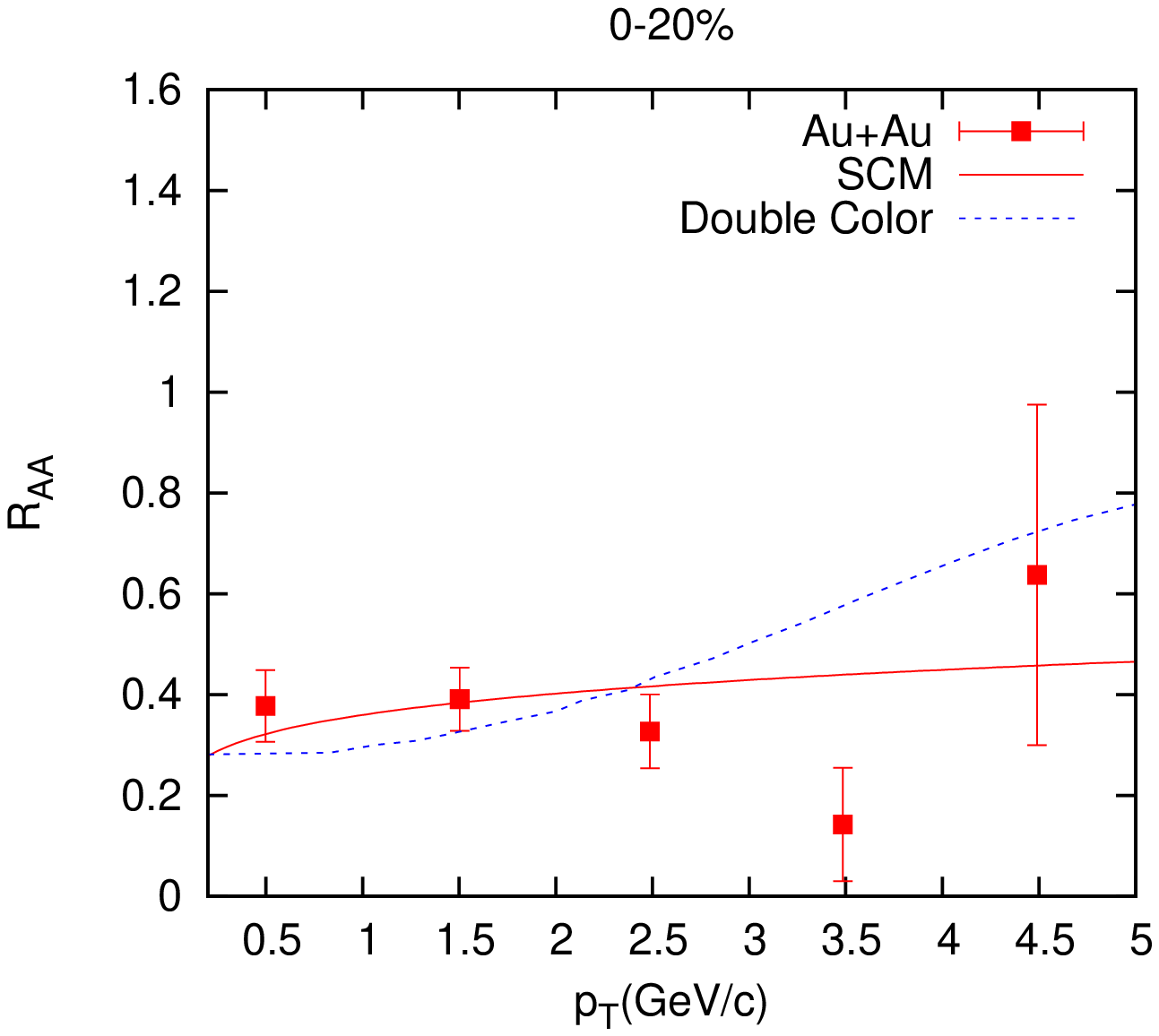}
  \end{minipage}}
  \caption{{ (a) Plot of the $R_{dAu}$ vs. $y$ for $J/\Psi$
production in $d+Au$ collisions at $\sqrt s_{NN} =200 GeV$. The data points are from \cite{adare4}. The solid
curves show the SCM-based results while the dotted curve depicts the Double Color Filter Approach \cite{kope}.
(b)Plot of the $R_{AA}$ for $J/\Psi$ production in$Au+Au$
collisions at $\sqrt s_{NN} =200 GeV$ as function of $p_T$. The data
points are from \cite{star}. The solid and dotted curves show respectively the SCM and the Double Color Filter-oriented \cite{kope}
results. } }
\end{figure}
\end{document}